# Spectroscopy of vanadium (III) doped gallium lanthanum sulphide chalcogenide glass

M. Hughes, H. Rutt and D. Hewak

*Optoelectronics Research Centre, University of Southampton, Southampton, SO17 1BJ, United Kingdom*

R. J. Curry

*Advanced Technology Institute, School of Electronics and Physical Sciences, University of Surrey, Guildford, GU2 7XH, United Kingdom*

Vanadium doped gallium lanthanum sulphide glass (V:GLS) displays three absorption bands at 580, 730 and 1155 nm identified by photoluminescence excitation measurements. Broad photoluminescence, with a full width half maximum (FWHM) of ~500 nm, is observed peaking at 1500 nm when exciting at 514, 808 and 1064 nm. The fluorescence lifetime and quantum efficiency at 300 K were measured to be 33.4 μs and 4% respectively. From the available spectroscopic data we propose the vanadium ions' valence to be 3+ and be in tetrahedral coordination The results indicate potential for development of a laser or optical amplifier based on V:GLS.

Chalcogenide glasses have low phonon energies due to the relatively large atomic mass of their constituent atoms. In particular gallium lanthanum sulphide (GLS) glass has a maximum phonon energy of 425 cm$^{-1}$, which gives it excellent infrared transmission up 9 μm.[1] This low maximum phonon energy enables emission from many transitions within active ion dopants, such as transition metals, which are weakly or not at all observed in other glasses (e.g. silica) due to their high phonon energies. In this paper we present spectroscopic data for vanadium doped GLS (V:GLS) glass. We assign a 3+ oxidation state to the vanadium ion and energy levels to the observed transitions by comparisons to previous work on the spectroscopic analysis of V$^{3+}$ in other hosts.

Samples of V:GLS were prepared by mixing 65% gallium sulphide (Ga$_x$S$_y$), 29.95% lanthanum sulphide (La$_2$S$_3$), 5% lanthanum oxide (La$_2$O$_3$) and 0.05% vanadium sulphide (V$_2$S$_3$) (% molar) in a dry-nitrogen purged glove box. Gallium and lanthanum sulphides were synthesised in-house from gallium metal (9N purity) and lanthanum fluoride (5N purity) precursors in a flowing H$_2$S gas system. Before sulphurisation lanthanum fluoride was purified and dehydrated in a dry-argon purged furnace at 1250 $^o$C for 36 hours to reduce OH$^-$ and transition metal impurities. The lanthanum oxide and vanadium sulphide were purchased commercially and used without further purification. The glass was melted in a dry-argon purged furnace at 1150 $^o$C for 24 hours before being quenched and annealed at 400 ºC for 12 hours.

Absorption spectra were taken on a Varian Cary 500 spectrophotometer over a range of 175-3300 nm with a resolution of ±0.1nm. Samples were cut and polished into 5mm and 0.5mm thick slabs which allowed reflection corrected absorption coefficient spectra to be calculated using equation 1.

$$\alpha_{rc}(\lambda) = \frac{A_{l_1}(\lambda) - A_{l_2}(\lambda)}{l_1 - l_2} \quad (1)$$

Where α$_{rc}$(λ) is the reflection corrected absorption coefficient spectrum and $A_{l_i}(\lambda)$ is the absorbance spectra of a sample of thickness $l_i$. Photoluminescence (PL) spectra were obtained by dispersing the fluorescence generated by laser sources at 514, 808 and 1064 nm in a Bentham TMc300 monochromator and detecting with a liquid nitrogen cooled InSb or InGaAs detector coupled with standard phase sensitive detection. All spectra were corrected for the system response. To obtain photoluminescence excitation (PLE) spectra a 1400 nm long pass filter was placed in front of an InGaAs detector to give an effective detection range of 1400-1700 nm. The excitation source used was a 250W quartz halogen white light source passed through a monochromator with a 5 nm bandwidth. The PLE spectra were corrected for the varying intensity of exciting light due to varying grating response and spectral output of the white light source by characterising the output of each grating with wavelength calibrated Newport 818-SL and 818-IG detectors and a Newport 1830-c optical power meter.

Fluorescence lifetime measurements were obtained using a 1064nm Nd:YAG laser modulated using a Gooch and Housego 80MHz acousto-optic modulator and the fluorescence was detected with a New Focus 2053 InGaS detector. The sample was placed in a Leybold AG helium gas closed cycle cryostat for cryogenic measurements.

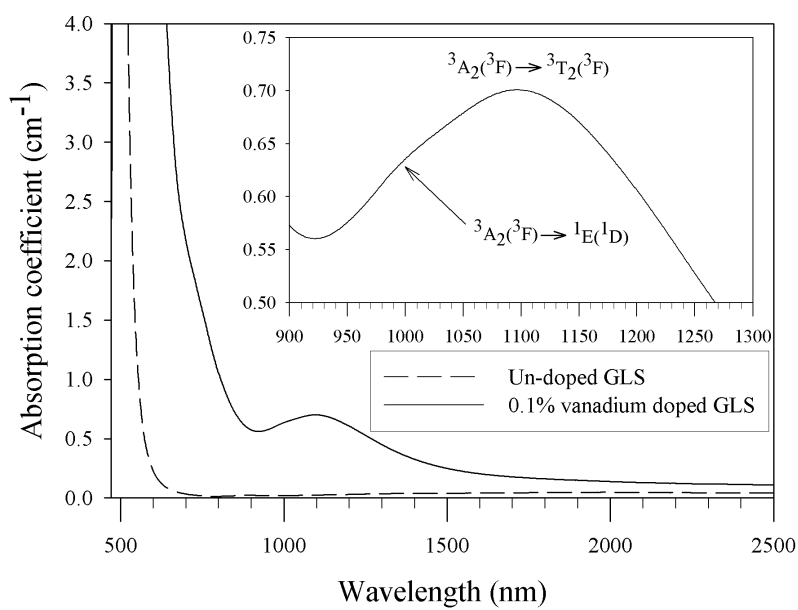

Figure 1 Reflection corrected absorption spectra of 0.1% vanadium doped GLS and un-doped GLS. Inset shows the peak at 1100 nm in detail

Figure 1 shows the absorption spectra of un-doped and 0.1% (molar) $V^{3+}$ doped GLS. The un-doped absorption spectrum is typical for that of GLS showing a strong electronic absorption edge at ~500 nm. Within the absorption spectrum of the 0.1% V:GLS sample a broad absorption centred at ~1100 nm and a shoulder at ~750 nm can be identified. The red shift of the absorption edge in the doped glass indicates that a third vanadium absorption band lies around 500nm. There is also evidence of a weak shoulder at ~1000nm which is attributed to a spin forbidden transition. No further absorption peaks were observed at wavelengths of up to 9 μm.

The PLE spectrum in figure 2 shows three broad Gaussian peaks located at 580, 730 and 1155 nm, which clarifies the identification of the two high energy absorption bands in figure 1. The slight red-shift of the strongest absorption band observed in the absorption and PLE spectra (centred at 1100 and 1155 nm respectively) can be reconciled by considering the overlap between the absorption and emission bands. When exciting at wavelengths longer than 1100 nm a greater proportion ions sited in a lower crystal field (and hence fluorescing at longer wavelengths) will be excited. This will shift the PLE peak to longer wavelengths as there would be less re-absorption of fluorescence.

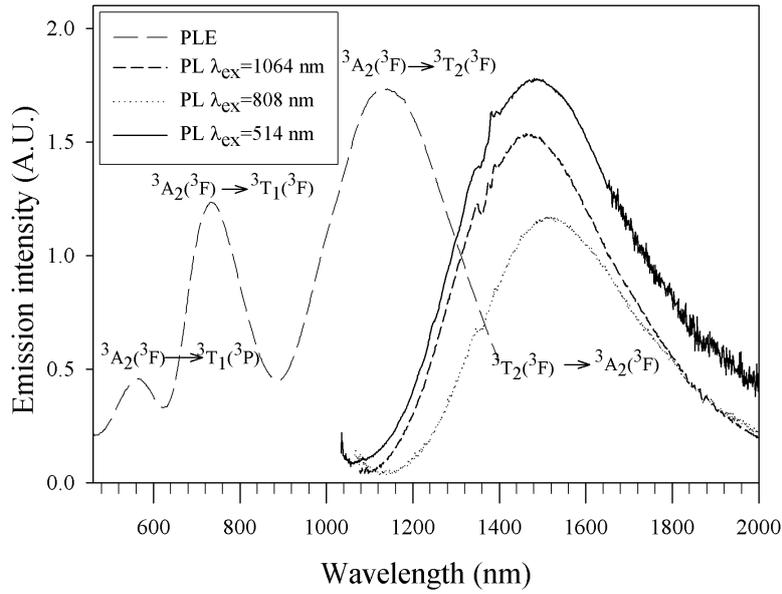

Figure 2. Excitation and photoluminescence spectra of 0.01% vanadium doped GLS.

Figure 2 shows the room temperature PL spectra taken using laser excitation sources at 514, 808 and 1064 nm which roughly equates to exciting into each of the three absorption bands identified in figures 1 and 2. The PL spectra taken at each excitation wavelength show similar characteristic spectra peaking at ~1500 nm with a full width half maximum (FWHM) of ~500 nm. This indicates that the three absorption bands belong to the same oxidation state rather than two or more oxidation states which is commonly observed in transition metal doped glasses[2] and crystals.[3] The broadness of the PL spectra indicates that the vanadium ion is in a low crystal field site. A result of this is that the $^3T_2(^3F)$ level, which is strongly dependent on crystal field strength, is the lowest energy level . Conversely, in a strong crystal field site the $^1E(^1D)$ level, which is almost independent of crystal field strength, is the lowest energy level. In this case characteristic narrow R-line emission should be observed as in $V^{3+}$ doped phosphate glass[4] and $V^{3+}$ doped corundum.[5]

The optical properties of V:GLS are similar to that of tetrahedrally co-ordinated $V^{3+}$ in other hosts. Tetrahedral $V^{3+}$ in yttrium aluminium garnet (YAG) has three absorption bands centred at 600, 800 and 1320 nm which are attributed to spin allowed transitions from the $^3A_2(^3F)$ ground state to the $^3T_1(^3P)$, $^3T_1(^3F)$ and $^3T_2(^3F)$ levels respectively, a weak and narrow absorption at 1140 nm is attributed to the spin forbidden transition $^3A_2(^3F)$ to $^1E(^1D)$.[3,6,7] Likewise, tetrahedral $V^{3+}$ in $LiAlO_2$, $LiGaO_2$ and $SrAl_2O_4$ has three absorption bands centred ~550, 850 and 1350 nm which are attributed to transitions from $^3A_2(^3F)$ to $^3T_1(^3P)$, $^3T_1(^3F)$ and $^3T_2(^3F)$ levels respectively.[8] Optical transitions associated with octahedral $V^{3+}$ tend to occur at higher energies than tetrahedral $V^{3+}$. For example the $^3T_1(^3F)$ to $^3T_2(^3F)$ and $^3T_1(^3P)$ transitions of octahedral $V^{3+}$:YAG occur at 600 nm and 425 nm respectively,[3,6,9] at 707 nm and 440 nm in zirconium fluoride glass[10] and at 724 nm and 459 nm in phosphate glass.[4]

Dopant ions in glasses are generally expected to substitute for network modifier cations.[13] The main network modifier in GLS is $La^{3+}$.[14] Using the approximate matrix elements of the Tanabe-Sugano model[15] for a $3d^2$ electronic configuration in octahedral coordination along with the absorption peaks at 1100 and 730 nm a crystal field strength (Dq/b) of 2.8 and a Racah c parameter of 1570 $cm^{-1}$ was calculated. This would mean the $V^{3+}$ ion is in a high field site which is inconsistent with the FWHM and lifetime of the PL. The same calculation for a $3d^2$ tetrahedral configuration gives Dq/b = 1.6 and c = 2460 $cm^{-1}$ which is a low field site and is consistent with the FWHM and lifetime of the PL and the spin forbidden absorption at 1000 nm. We therefore propose that $V^{3+}$ is tetrahedrally co-ordinated and attribute absorption peaks at 580, 730 and 1100 nm to transitions from the $^3A_2(^3F)$ ground state to the $^3T_1(^3P)$, $^3T_1(^3F)$ and $^3T_2(^3F)$ levels; the weak shoulder at 1000 nm to the $^3A_2(^3F)$ to $^1E(^1D)$ transition; and the PL peaking at 1500 nm to the $^3T_2(^3F)$ to $^3A_2(^3F)$ transition. It is also noted that there is no significant change in absorption, photoluminescence and lifetime of $V^{3+}$ in the host gallium lanthanum oxy-sulphide (GLSO) which contains ~15% (molar) oxygen.

The fluorescence decay profiles were found to be non-exponential but could be accurately described using a stretched exponential function: $I(t)=I_0 \exp[-(t/\tau)^p]$. Where $I_0$ is the initial fluorescence intensity, $\tau$ is the fluorescence lifetime and P is the stretch factor. The stretched exponential function is widely known to fit many different relaxation processes including: charge relaxation, magnetic susceptibility relaxation and fluorescence decay in many amorphous materials[16,17] and crystalline solids.[18] Stretched exponential decay behaviour can be explained by the random network structure of glasses.[19]

Using this method to fit the fluorescence decay profiles shown in Figure 3 lifetimes of 54.5 μs and 33.3 μs are obtained at 6.5 K and 300 K respectively. This increase in lifetime with reduction in temperature is attributed to the depopulation of higher phonon levels at lower temperatures which reduces the probability of non-radiative decay. For comparison the 300 K lifetimes for $V^{3+}$ doped YAG, $LiAlO_2$ and $LiGaO_2$ are 22 ns [3], 0.5 μs [20] and 11 μs [20] respectively.

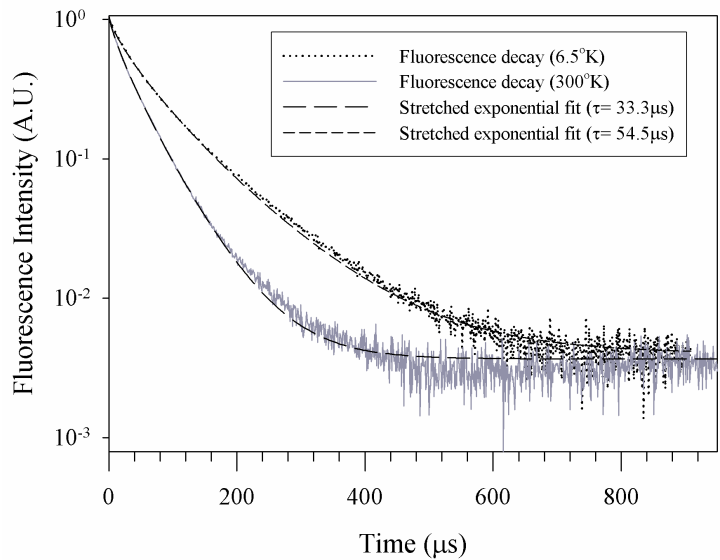

Figure 3 Fluorescence decay of vanadium doped GLS at 6.5K and 300K fitted with stretched exponentials.

Quantum efficiency (QE) measurements were obtained by placing the sample in an integrating sphere and taking spectra of the fluorescence and of the 1064 nm exciting laser line, with and without the sample present. A "photons out/photons in" method similar to that described elsewhere was used to calculate the QE.[21,22] The photons out were calculated from the area under the fluorescence spectra and the photons in were calculated from the difference in the area of the laser line spectra with and without the sample present. The QE of 0.1% V:GLS was calculated to be 4.2% .

The formula of McCumber[23] in equation 3 is used to calculate the peak emission cross section. Where $\lambda_0$ is the peak

$$\sigma_{em} = \sqrt{\frac{\ln 2}{\pi}} \frac{A}{4\pi c n^2} \frac{\lambda_0^4}{\Delta\lambda} \tag{2}$$

fluorescence wavelength, $\Delta\lambda$ is the FWHM, n is the refractive index, c is the speed of light and A is the Einstein coefficient calculated from: $QE/\tau = 1260 s^{-1}$. The calculation for V:GLS is given in table I. To date there has been no demonstration of lasing from $V^{3+}$ in part due to the small radiative lifetimes mentioned above. Table I compares the spectroscopic parameters from this work with those of other relevant laser materials doped with $Cr^{4+}$ which is isoelectric to $V^{3+}$, $V^{2+}$ and the commercially successful Ti:Sapphire. Comparisons indicate that the lifetime of V:GLS is comparable or better than existing doped laser hosts. Though the QE and emission cross-section do not compare favourably, the ability to form optical fibres from this material may overcome potential heat dissipation problems caused by the low QE (due to the large surface area to volume ratio of optical fibres). Additionally, the high pump beam confinement that can also be achieved in a fibre could compensate for the low emission cross section.

Table I Overview of the spectroscopic parameters for various laser materials compared to V:GLS

| Ion:Host | $\tau(\mu s)$ | $\eta_{QE}$ (%) | $\sigma_{em}$ ($10^{-19}cm^2$) | $\sigma_{em}.\tau$ ($10^{-24}scm^2$) | Ref. |
|---|---|---|---|---|---|
| $V^{3+}$:GLS | 33 | 4 | 0.03 | 0.1 | This work |
| $Cr^{4+}$:Y$_2$SiO$_5$ | 1.3 | <10 | 2.0 | 0.26 | 24 |
| $Cr^{4+}$:YAl$_5$O$_{12}$ | 4.1 | 22 | 3.3 | 1.35 | 25 |
| $V^{2+}$:MgF$_2$ | 40 | | 0.08 | 0.32 | 26 |
| $Ti^{3+}$:Al$_2$O$_3$ | 3.1 | 100 | 4.5 | 1.40 | 27 |

In summary, we have demonstrated strong, broad, room temperature emission peaking at ~1500 nm from $V^{3+}$ doped GLS glass when excited by laser sources into each of the three absorption bands at 580, 730 and 1100 nm. The room temperature emission lifetime was measured to be 33.3μs. We note that this system may have potential for developing a low-cost V:GLS tuneable laser or optical amplifier, with an optical fibre wave-guiding structure being best possibility for achieving this.